\documentclass[smallextended,natbib,runningheads]{svjour3}
\journalname{Annals of the Institute of Statistical Mathematics}
\smartqed  
\usepackage{placeins}
\usepackage{graphicx}
\usepackage{amssymb}	
\usepackage{amsmath}	
\usepackage{multirow} 			
\usepackage{soul} 			
\usepackage{diagbox}		
\usepackage{commath}    
\usepackage[mathscr]{euscript} 
\usepackage{hyperref} 
\usepackage{xcolor} 
\newcommand{\e}{\mathrm{I \! E}} 
\newcommand{\re}{\mathrm{I \! R}} 
\newcommand{\na}{\mathrm{I \! N}} 
\newcommand{\var}{\mathrm{Var}} 
\newcommand{\cov}{\mathrm{Cov}} 
\newtheorem{prop}{Proposition} 

\newcommand{\vb}{\textbf{b}}

\newcommand{\vbeta}{\pmb{\beta}}

\newcommand{\vomega}{\pmb{\omega}}

\newcommand{\vgamma}{\pmb{\gamma}}

\newcommand{\vxi}{\pmb{\xi}}

\begin{document}

\title{Wavelet estimation of the dimensionality of curve time series\thanks{The first author acknowledges FAPESP Grant 2016/24469-6 . The second  author acknowledges FAPESP Grants 2013/00506-1 and 2018/04654-9 and CNPq Grant 309230/2017-9.
}
}
\subtitle{ }


\author{Rodney V. Fonseca \and  Alu\'{i}sio Pinheiro 
}


\institute{Rodney V. Fonseca (Corresponding author) \at
              Department of Statistics, University of Campinas, Brazil \\
              \email{rodneyfv@gmail.com}           
           \and
           Alu\'{i}sio Pinheiro \at
               Department of Statistics, University of Campinas, Brazil 
}

\maketitle

\begin{abstract}
Functional data analysis is ubiquitous in most areas of sciences and engineering. Several paradigms are proposed to deal with the dimensionality problem which is inherent to this type of data. Sparseness, penalization, thresholding, among other principles, have been used to tackle this issue. We discuss here a solution based on a finite-dimensional functional space. We employ wavelet representation of the functionals to estimate this finite dimension, and successfully model a time series of curves. The proposed method is shown to have nice asymptotic properties. Moreover, the wavelet representation permits the use of several bootstrap procedures, and it results in faster computing algorithms. Besides the theoretical and computational properties, some simulation studies and an application to real data are provided.

\vspace*{.5cm}


\end{abstract}

\section{Introduction}
\label{intro}

Many phenomena, natural or anthropogenic, can be appropriately modeled by a function on a suitable domain. Examples on the literature have been around for several decades, but the last three have made them ubiquitous in most areas of science and engineering such as, but not limited to, physics, astronomy, chemistry, genetics, biology, macroeconomics, medicine, energy, microeconomics, finance, digital communication, medical imaging, insurance, oceanography, psychology and anthropology. The underlying stochastic structure of these high-dimensional data can be understood as a technical tool towards reproductibility/repeatability or inherent to the problem under study. Either way, a precise apportionment of deterministic and random components is paramount. Examples of relevant data sets and areas as well as paradigms for the statistical analysis of functional data ca be found in \cite{ramsay2005functional} and \cite{morettin2017wavelets}. 

Some features are found in specific problems, and should be dealt with accordingly. For instance, intraday and/or inter-day dependences is common in financial functional series \citep{MR3601312,MR3053624,MR2848514}. Aggregate data may be useful for energy \citep{dias2013hierarchical, dias2015aggregated}, market shares \citep{MR3268395}, demand/supply studies  \citep{MR3572849}, and many others \citep{MR3553239, MR3496450, MR3496460}. Besides these particular characteristics, the proposed model must deal with a very common and basic property: the dimension of the functional space and its impact on the proposed solution. Functional data analysis poses serious hindrances to parametric models. Some Bayesian proposals that deal with this dimensionality issues can be found at \cite{MR3580124}, \cite{MR3620735}, and \cite{MR3572849}.

When pursuing nonparametric analysis of high-dimensional and functional data, the model dimension must be limited as well. Successful solutions are found  through sparseness \citep{MR3804247, MR3463795, MR3533642, MR3463793, MR3656031, MR3800150, MR3640189, MR3682621}, principal component analysis (Mousavi and S{\o}rensen 2018); \citep{MR3533642, MR3732338, MR3713402, MR3607885, MR3535037, MR3553239}, thresholding \citep{MR3740728, MR3506434, MR3727508, MR3613398, MR3681117,roislien2013feature,salvatore2016exploring,johnstone2009consistency}, penalizing procedures \citep{MR3740728, MR3646200, MR3727508, MR3396986, MR3640189, MR3627011}, sufficiency \citep{MR3732339, MR3662448}, and others \citep{MR3580073, MR3656031, MR3682621, MR3611771}.  

Here we follow the set-up studied by \cite{bathia2010identifying}. The idea is to model the functional space as driven by a finite-dimensional basis plus some noise term. This can be applied to a time series composed of curves. This problem was also studied by \cite{hall2006assessing}, who proposes a way to estimate the number of components of the functional's covariance using the assumption that the noise vanishes as the sample size increases. The methodology proposed by \cite{bathia2010identifying} does not need such assumption, exploring instead the dynamic structure of the observed curves. Eigenfunctions are used to represent the curves and bootstrap resampling is proposed to sequentially estimate the finite functional dimension.
The methods of the latter were used, for instance, by \cite{horta2018dynamics} to compute the dimension of time series density functions of stock indexes for prediction purposes. 

In this work we employ wavelet bases to build the curves, and estimate the functional dimension. This novelty on the basis allows us to propose a series of bootstrapping procedures besides the original one by \cite{bathia2010identifying}. Similar asymptotic properties are attained. Moreover, computational and mathematical advantages are discussed. We also prove that the estimation procedure may be used for aggregate data.

The text is organized as follows. In Section \ref{Sec:methodology} we discuss the idea of finite functional dimension. In Section \ref{Sec:WaveletFuncDim} we present the proposed wavelet solution for the estimation of the functional dimension. Two cases of particular interest are discussed in Section 
\ref{Sec:cases_of_interest}. We then present the theoretic results for the proposed algorithms in Section \ref{Sec:theoretical}. Simulation studies and an application to real data are presented in Sections \ref{Sec:simulations} and \ref{Sec:Applic}, respectively. A discussion and final remarks can be found in Section \ref{Sec:discussion}.

\section{Functional dimension estimation}
\label{Sec:methodology}

In what follows we shall describe the problem of estimating the finite dimension of curve time series \citep{bathia2010identifying}. Consider random functions $Y_1,Y_2,\ldots$ in a Hilbert space $L^2=L^2(I)$ of square integrable functions defined in a compact $I\subset\re$, with inner product $\langle Y,X\rangle=\int_I Y(x)X(x)dx$, $\forall Y,X\in L^2$. These curves usually are not perfectly observed, being subject to errors of numerical or experimental nature, for example. This means that in practice we do not know the curves of interest $X_t$, $t=1,\ldots,n$, but we might have a sample of estimates $Y_1,\ldots,Y_n$ obtained after applying some smoothing method to the data at hand. The observed curves $Y_t$ are taken as satisfying
\begin{align}
Y_t(x) = X_t(x) + \varepsilon_t(x), \quad x\in I,t=1,\ldots,n,
\label{E:bathia_model}
\end{align}
where $X_t$ and $\varepsilon_t$ are not observed and $\varepsilon_t$ is supposed to be a noise, in the sense that
\begin{enumerate}
\item $\e[\varepsilon_t(x)]=0$, $\forall t$ and $\forall x\in I$,
\item $\mathrm{Cov}(\varepsilon_t(x),\varepsilon_{t+k}(y))=0$, $\forall x,y\in I$ when $k\neq 0$,
\item $\mathrm{Cov}(X_t(x),\varepsilon_s(y))=0$, $\forall x,y\in I$ and $\forall t,s$.
\end{enumerate}
With these conditions, the error of estimating $X_t$ is intrinsic to time $t$ and exogenous with respect to $X_t$. We assume that $X_1,X_2,\ldots$ are stationary, such that
\begin{align*}
\mu(x) = \e[X_t(x)] \quad\text{and}\quad M_k(x,y) = \mathrm{Cov}(X_t(x),X_{t+k}(y)),
\end{align*}
do not depend on $t$. Under the assumption that $X_t$ is a second order-processes it admits the Karhunen-Loève expansion \citep{bosq2000linear}, and \cite{bathia2010identifying} consider that the process of interest has dimension $d\in\na$, such that it holds the following spectral decomposition:
\begin{align}
X_t(x) = \mu(x) + \sum_{j=1}^{d}\xi_{tj}\varphi_{j}(x), \quad \forall t
\label{E:ProcessSpecDecomp}
\end{align}
where $\xi_{tj}=\langle X_t-\mu,\varphi_j\rangle$ is a zero-mean random variable and $\varphi_1,\ldots,\varphi_d$ compose an orthonormal set in $L^2$ of eigenfunctions of the positive kernel
\begin{align*}
K(x,y) = \sum_{k=1}^{p}\int M_k(x,z)M_k(y,z)dz.
\end{align*}
Under the representation of $X_t$ given by (\ref{E:ProcessSpecDecomp}), the dynamics of $Y_1,Y_2,\ldots$ is captured through the $d$-dimensional time series $\pmb{\xi}_t=(\xi_{t1},\ldots,\xi_{td})^{\top}$. We consider the observed kernel
\begin{align*}
\hat{K}(x,y)=\sum_{k=1}^{p}\int \hat{M}_k(x,z)\hat{M}_k(y,z)dz,
\end{align*}
where $p$ is fixed and
\begin{align*}
\hat{M}_k(x,y) = \frac{1}{n-p}\sum_{t=1}^{n-p}(Y_t(x)-\bar{Y}(x))(Y_{t+k}(y)-\bar{Y}(y)),
\end{align*}
with $\bar{Y}(x)=\sum_{t=1}^{n}Y_t(x)/n$.

The maximum lag $p$ in practice can be taken as a small integer positive value \citep{bathia2010identifying}. The authors idea in identifying $d$ is to obtain eigenfunctions of $\hat{K}(x,y)$ through eigenvectors and eigenvalues of a finite dimension matrix whose elements are computed from inner products involving $Y_t$ and $\bar{Y}$. We employ in this paper wavelet representation to perform the eigenanalysis of $\hat{K}(x,y)$. In Section \ref{Sec:WaveletFuncDim} we briefly introduce wavelet methods and present the proposed wavelet procedure for dimension estimation.

\section{Wavelet based functional dimension}
\label{Sec:WaveletFuncDim}

Wavelet methods are useful to approximate functions in $L^2(\re)$ with a basis $\{\psi_{j,k};j,k\in\mathbb{Z}\}$ whose elements are obtained through translation and dilation operations of a function $\psi$ (the wavelet function)
\begin{align*}
\psi_{j,k}(t) = 2^{j/2}\psi(2^{j/2}t - k), \quad j,k\in\mathbb{Z}.
\end{align*}
\cite{meyer1985principe} shows that any function $f\in L^2(\re)$ can be written (in $L^2$ sense) as
\begin{align}
\label{E:serie_psi_f}
f(t) = \sum_{j\in\mathbb{Z}}\sum_{k\in\mathbb{Z}}c_{j,k}\psi_{j,k}(t),
\end{align}
whose coefficients are given by $c_{j,k} = \langle f,\psi_{j,k}\rangle$.

Orthonormal wavelet basis can be constructed by means of a Multiresolution Analysis (MRA), which is a tool presented by \cite{mallat1989theory} that consists of a nested sequence of closed subspaces $\{V_n,n\in\mathbb{Z}\}$ in $L^2(\re)$ satisfying:
\begin{enumerate}
\item $\cdots\subset V_{-2}\subset V_{-1}\subset V_{0}\subset V_{1}\subset V_{2}\subset\cdots$;
\item $\cap_{n}V_n = \{0\}$ and $\overline{\cup_{n}V_n}=L^2(\re)$;
\item the subspaces $V_n$ are self-similar, in the sense that $f(2^jx)\in V_j \Leftrightarrow f(x)\in V_0$;
\item exists a function $\phi\in V_0$ that composes a orthogonal basis of $V_0$ in the following way:
\begin{align*}
V_0 = \left\{ f; f(x)=\sum_{k\in\mathbb{Z}}c_k\phi(x-k)\right\}.
\end{align*}
\end{enumerate}

In the literature, $\phi$ is known as a scale function and the subspaces $V_j$ can be seen as resolution levels by which we can approximate a function on $L^2(\re)$. Using the self-similarity of the subspaces $V_j$, we have $\{\phi_{j,k}(x)=2^{j/2}\phi(2^{j}x-k),k\in\mathbb{Z}\}$ is a basis for $V_j$. \cite{mallat1989theory} shows that any function $f\in L^2(\re)$ can be approximated in $V_j$ by
\begin{align*}
P_j f(x) = \sum_{k\in\mathbb{Z}}\langle f,\phi_{j,k} \rangle \phi_{j,k}(x),
\end{align*}
where $P_j f$ here denotes the orthogonal projection of $f$ in $V_j$. Based on properties 1 and 2 of the MRA we have $\lim_{j\rightarrow\infty}P_j f(x) = f(x)$ and  $\lim_{j\rightarrow -\infty}P_j f(x) = 0$, i.e., higher resolutions provide better approximations to $f(x)$ whereas, the lower the resolution is, the closer to zero is the approximation. The rate of this approximation can be evaluated when $f$ belongs to certain functional spaces, like Sobolev and Besov spaces \citep{hardle1998wavelets}.

The detail obtained after passing from a resolution $j$ to $j+1$ can be analyzed considering the orthogonal complement of $V_j$ in $V_{j+1}$, which is denoted by $W_j$. Hence, $V_{j+1}=V_j\oplus W_j$, which gives
\begin{align*}
V_j=\bigoplus_{k<j}W_k \quad\text{and}\quad L^2(\re) = \bigoplus_{k\in\mathbb{Z}}W_k.
\end{align*}
\cite{mallat1989theory} shows that the wavelet function $\psi$ has the property that $\{\psi_{j,k}(x)\!=\!2^{j/2}\psi(2^{j}x-k),k\in\mathbb{Z}\}$ is an orthonormal basis of $W_j$ and $\{\psi_{j,k}(x)=2^{j/2}\psi(2^{j}x-k);k\in\mathbb{Z},j\in\mathbb{Z}\}$ is a basis of $L^2(\re)$. The wavelet function $\psi$ can be used in a series representation of $f\in L^2(\re)$ like Equation (\ref{E:serie_psi_f}), and from the MRA we have that $f$ can also be represented as
\begin{align}
\label{E:series_psi_phi_f}
f(x) = \sum_{k\in\mathbb{Z}}\langle f,\phi_{j_0,k} \rangle \phi_{j_0,k}(x) + \sum_{j\geq j_0}\sum_{k\in\mathbb{Z}}\langle f,\psi_{j,k} \rangle \psi_{j,k}(x),
\end{align} 
where the first series is the projection of $f$ in resolution $j_0$, $\langle f,\phi_{j_0,k} \rangle$ being called as an approximation coefficient, and the second series contains the details corresponding to resolutions greater or equal to $j_0$, with $\langle f,\psi_{j,k} \rangle$ being called a detail coefficient. A widely used system is the Daubechies wavelets, which have compact support and nice properties regarding function regularity. We denote DAUB$N$ as a Daubechies wavelet with $N$ null moments. The case $N=1$ corresponds to the famous Haar wavelet \citep{vidakovic2009statistical}. Since $V_0\subset V_1$
\begin{align}
\phi(x) = \sum_{k\in\mathbb{Z}}h_k\sqrt{2}\phi(2x-k),
\label{E:scaling_equation}
\end{align}
where the vector of coefficients $\mathbf{h}=\{h_k,k\in\mathbb{Z}\}$ is known as the wavelet filter. Equation (\ref{E:scaling_equation}) is known as scaling equation and is important in the computation of wavelets \citep{vidakovic2009statistical}. In addition, since $W_0\subset V_1$, we have that
\begin{align}
\psi(x) = \sum_{k\in\mathbb{Z}}g_k\sqrt{2}\phi(2x-k),
\label{E:psi_equation}
\end{align}
for some coefficients $\mathbf{g}=\{g_k,k\in\mathbb{Z}\}$. It is possible to show that $g_n=(-1)^n h_{1-n}$, which is called the quadrature mirror relation. Other properties of the coefficients $h_k$ and $g_k$ can be found in \cite{vidakovic2009statistical}. These coefficients play an important role on the computation of wavelet and scale functions. For instance, they are used to compute the discrete wavelet transformation with a cascade algorithm that uses $\mathbf{h}$ and $\mathbf{g}$ as filters of convolution operators.

Our idea is to employ wavelet decompositions of the observed functionals $Y_t$ to estimate the dimension of the process that generates these curves. To that end, we use $Y_t$ evaluated in a grid of points selected in a appropriate way as follows. For notational convenience, consider
\begin{align*}
Y_t(x) = \sum_{j}a_{j}^t\phi_{j}(x),
\end{align*}
where $a_{j}^t$ represents both wavelet and approximation coefficients and $\phi_{j}$ represents both scale and wavelet functions. Hence
\begin{align*}
Y_t(x)-\bar{Y}(x) = \sum_{j}(a_{j}^t - \bar{a}_{j})\phi_{j}(x)
= \sum_{j}c_{j}^t\phi_{j}(x),
\end{align*}
where $\bar{a}_{j}=n^{-1}\sum_{t=1}^{n}a_{j}^t$. Therefore, we obtain
\begin{align*}
\hat{M}_k(x,y)=\frac{1}{n-p}\sum_{t=1}^{n-p}\sum_{j}\sum_{l}c_{j}^t\phi_{j}(x)\phi_{l}(y)c_{l}^{t+k}
\end{align*}
and using that the functions $\phi_j(x)$ form an orthonormal system, we get
\begin{align*}
\hat{K}(x,y) = \frac{1}{(n-p)^2}\sum_{k=1}^{p}\sum_{t=1}^{n-p}\sum_{s=1}^{n-p}\sum_{j,j',l}c_{j}^t c_{j'}^s c_{l}^{t+k} c_{l}^{s+k} \phi_{j}(x)\phi_{j'}(y).
\end{align*}

Our objective is to find eigenfunctions of the operator $\hat{K}(x,y)$. Considering a candidate eigenfunction $h_m$, its wavelet representation is given by
\begin{align}
h_m(x) = \sum_{j''}b^m_{j''}\phi_{j''}(x) = \Phi(x)^{\top}\vb^m,
\label{E:psi_m}
\end{align}
where $\Phi(x) = (\phi_1(x),\ldots,\phi_J(x))^{\top}$ and $\vb^m = (b_1^m,\ldots,b_J^m)^{\top}$. Note that we define $J$ as the number of terms in the wavelet decompositions. Thus, considering that the same basis are being used in all decompositions, the indexes $j$, $j'$ and $j''$ also vary in $\{1,\ldots,J\}$. Then
\begin{align}
\int \hat{K}(x,y)h_m(y)dy &= \sum_{j}\left(\frac{1}{(n-p)^2}\sum_{k=1}^{p}\sum_{t=1}^{n-p}\sum_{s=1}^{n-p}\sum_{j',l}c_{j}^t c_{j'}^s c_{l}^{t+k} c_{l}^{s+k} b^m_{j'}\right) \phi_{j}(x) \nonumber \\
&=\sum_{j} (D_{j}\vb^m) \phi_{j}(x) = \Phi(x)^{\top}(D\vb^m),
\label{E:M_psi_m}
\end{align}
where $D_{j}$ represents a $1\times J$ vector which is the $j$-th row of the $J\times J$ matrix $D$, whose $(j,j')$ element is
\begin{align*}
D_{j,j'} = \frac{1}{(n-p)^2}\sum_{k=1}^{p}\sum_{t=1}^{n-p}\sum_{s=1}^{n-p}\sum_{l}c_{j}^t c_{j'}^s c_{l}^{t+k} c_{l}^{s+k}.
\end{align*}
This matrix can also be obtained in the following way. Consider the $J\times n$ matrix $\mathcal{C}$ whose $t$-th column contains $J$ coefficients $c_{j}^{t}$, then letting $\mathcal{C}_{J\times(k_1:k_2)}$ be a submatrix obtained selecting from the $k_1$-th until the $k_2$-th column of $\mathcal{C}$, $k_1<k_2$, we have
\begin{align*}
D\! =\! \frac{1}{(n-p)^2}\mathcal{C}_{J\times(1:n-p)}\!\! \left( \sum_{k=1}^p(\mathcal{C}_{J\times(k+1:n-p+k)})^{\top}\mathcal{C}_{J\times(k+1:n-p+k)}\!\right)\!\!(\mathcal{C}_{J\times(1:n-p)})^{\top}.
\end{align*}
Therefore, from (\ref{E:psi_m}) and (\ref{E:M_psi_m}), our goal is to find $\vb^m$ such that $\Phi(x)^{\top}(D\vb^m) = \lambda_m \Phi(x)^{\top}\vb^m$ for some constant $\lambda_m$ and $\forall x\in I$, i.e., we wish to solve for $\vb^m$ the system
$$
(D\vb^m) = \lambda_m \vb^m,
$$
i.e., taking $\vb^m$ as an eigenvector of $D$, with $\lambda_m$ being its associated eigenvalue. Thus, letting $\vb^1,\ldots,\vb^{\hat{d}}$ be the eigenvectors of $D$ associated to its $\hat{d}$ largest eigenvalues, we have that $h_1,\ldots,h_{\hat{d}}$ as in Equation (\ref{E:psi_m}) are eigenfunctions of the operator $\hat{K}$. It is worth mentioning that this procedure resembles the functional PCA \cite[p. 162]{ramsay2005functional}, with the difference that instead of the matrix $D$ we would consider for the latter the $J\times J$ matrix $n^{-1}\mathcal{C}\mathcal{C}^{\top}$.

Since the decomposition coefficients of the $h_m$'s are orthonormal, we have that $\{h_1(\cdot),\ldots,h_{\hat{d}}(\cdot)\}$ forms an orthonormal system in $L^2$, the estimate of the functional of interest being thus
\begin{align}
\hat{Y}_t(x) = \bar{Y}(x) + \sum_{l=1}^{\hat{d}}\hat{\eta}_{tl}h_{l}(x),
\label{E:estim_functional}
\end{align}
where $\hat{\eta}_{tj}=\langle Y_t-\bar{Y},h_j\rangle$. Hence, it follows from (\ref{E:estim_functional}) that $\hat{\eta}_{tl}=\sum_{j}c_{j}^tb^l_{j}$, with the dynamics of $Y_t$ being modeled through the multivariate time series $\hat{\pmb{\eta}}_t = (\hat{\eta}_{t1},\ldots,\hat{\eta}_{t\hat{d}})'$. 

We compare four bootstrap procedures for the estimation of $d$. The first is described by \cite{bathia2010identifying}. Given the eigenvalues $\lambda_1\geq\lambda_2\geq\ldots\geq0$, tests of the null hypothesis $\lambda_{d_0+1}=0$ are performed sequentially until the first $\lambda_{d_0 + 1}$ which significantly equals zero is found. In this case, the estimated dimension is taken as $d_0$. For instance, suppose that $t\in\{1,\ldots,n\}$ and that we want to test $H_0:\lambda_{d_0+1}=0$ against $H_1:\lambda_{d_0+1}>0$ for some positive integer $d_0$. Then we construct a functional imposing the restriction of $H_0$:
\begin{align*}
\tilde{Y}_t(x) = \bar{Y}(x) + \sum_{l=1}^{d_0}\hat{\eta}_{tl}h_{l}(x)
= \sum_{j}\left\{ \bar{a}_{j} + \sum_{l=1}^{d_0}\hat{\eta}_{tl}b^l_{j} \right\}\phi_{j}(x),
\end{align*}
for which we already have the wavelet decomposition of $\tilde{Y}_t(x)$. Then we obtain the residuals $\hat{\epsilon}_t(x) = Y_t(x) - \tilde{Y}_t(x)$ and perform the following steps:
\begin{enumerate}
\item for each $t=1,\ldots,n$, randomly select (with replacement) a residual $\epsilon^{\text{b}}_t(x)$ from $\{\hat{\epsilon}_1(x),\ldots,\hat{\epsilon}_n(x)\}$ and take $Y_t^{\text{b}}(x) = \tilde{Y}_t(x) + \epsilon^{\text{b}}_t(x)$;
\item obtain for the bootstrap sample $Y_1^{\text{b}}(x),\ldots,Y_n^{\text{b}}(x)$ the matrix $D$, and compute its $(d_0+1)$-th largest eigenvalue $\lambda_{d_0+1}^{\text{b}}$;
\item repeat steps 1 and 2 a large number of times, say $B$, then compute the bootstrap p-value $p_{\text{boot}} = \#\{\hat{\lambda}_{d_0+1}<\lambda_{d_0+1}^{\text{b}}\}/(B+1)$, where $\hat{\lambda}_{d_0+1}$ is the $(d_0+1)$-th largest eigenvalue obtained for $\hat{Y}$. Reject $H_0$ if $p_{\text{boot}}$ is lower than some previously specified significance value.
\end{enumerate}
From the wavelet decompositions of $\bar{Y}$ and $h_l$, $l=1,\ldots,d_0$, we have
\begin{align*}
Y^{\text{b}}_t(x) &= \tilde{Y}_t(x) + (Y_t(x) - \tilde{Y}_t(x))^{\text{b}} 
&= \sum_{j}\left\{ a_{j}^{t{\text{b}}} + \sum_{l=1}^{d_0}(\hat{\eta}_{tl} - \hat{\eta}_{tl}^{\text{b}})b^l_{j} \right\}\phi_{j}(x),
\end{align*}
where the superscript ${\text{b}}$ indicates the bootstrapped terms. Hence, the resampling of step 1 can be performed directly on the coefficients $a_j^t$ and $\hat{\eta}_{tl}$, which reduces the computation time of the bootstrap procedure.

Three other bootstrap procedures follow the same algorithm with some modifications. Initially, note that in the previous bootstrap test the wavelet decomposition of the observed functionals is obtained without thresholding and then we apply the bootstrap procedure to test the eigenvalues of the corresponding matrices $D$. Taking this into account, in the second bootstrap procedure we perform the same steps as above with the only difference that a hard thresholding is applied for the observed functionals coefficients before computing $D$. In the third procedure we apply a hard thresholding (indicated by the index $thr$) to $\bar{Y}$ and $h_l$ in Equation (\ref{E:estim_functional}), such that
\begin{align*}
Y_t(x) = \bar{Y}(x) + \sum_{l=1}^{\hat{d}}\hat{\eta}_{tl}h_{l}(x) + \hat{\epsilon}_t(x)
= \bar{Y}^{thr}(x) + \sum_{l=1}^{\hat{d}}\hat{\eta}_{tl}h_{l}^{thr}(x) + \hat{\epsilon}^{thr}_t(x),
\end{align*}
where
\begin{align*}
\hat{\epsilon}^{thr}_t(x) = \hat{\epsilon}_t(x) + (\bar{Y}(x) - \bar{Y}^{thr}(x)) + \sum_{l=1}^{\hat{d}}\hat{\eta}_{tl}(h_{l}(x) - h_{l}^{thr}(x)).
\end{align*}
Then, we apply the bootstrap procedure with $\hat{\epsilon}^{thr}_t(x)$ instead of $\hat{\epsilon}_t(x)$ and $\tilde{Y}_t(x)$ formed of $\bar{Y}^{thr}$ and $h_l^{thr}$. Hence, we have that the bootstrap functional in this case is
\begin{align*}
Y^{\text{b}}_t(x) = \sum_j\left\{ a_j^{t{\text{b}}} + \sum_{l=1}^{d_0}[\hat{\eta}_{tl} - \hat{\eta}_{tl}^{\text{b}}](b_j^l)^{thr}\right\}\phi_j(x),
\end{align*}
which is similar to the previous bootstrap method, using the thresholded wavelet coefficient $(b_j^l)^{thr}$ and $a_j^{t{\text{b}}}$ instead of $b_j^l$ and $a_j^{t}$. The last procedure we consider is based on the wavestrapping technique proposed by \cite{percival2000wavestrapping}, where for each $t$, a residual $\epsilon_t^{\text{b}}(x)$ is randomly selected from $\{\hat{\epsilon}_1(x),\ldots$, $\hat{\epsilon}_n(x)\}$, then its wavelet coefficients are resampled (with replacement) inside each detail level to obtain the coefficients of a new bootstrap residual, which is used to form the bootstrap functional $\tilde{Y}_t(x)$. An advantage of the wavestrapping over the first three bootstrap methods is that it has a much larger number of possible residuals, since it is based not only on random selection from $n$ elements, but also considers resampling from their wavelet coefficients to obtain random samples.

\section{Two cases of practical interest}
\label{Sec:cases_of_interest}

The proposed wavelet method of functional dimension estimation can be applied on a variety of cases of time series curves. In this section we highlight two such cases.

\subsection{Functional data aggregation}

There is considerable attention on the literature to investigate situations where analyzing curves individually is not possible or very costly, requiring an analysis based on aggregate curves, as described in the references from Section 1. Moreover, aggregate time series are also well described in the literature, as can be seen in \cite[Chapter~20]{wei2006time}, since this kind of data is often found, as happens with economic data \citep{abraham1982temporal}, for example. In this paper we consider aggregate data to identify the dimension of a functional time series. 

Suppose that the observed curves as defined by (\ref{E:bathia_model}) are not originally aggregates, but the number of observations for each time is not large. We then take linear combinations of $Y_t(\cdot)$ for a fixed number of successive $t$'s as observed functions before applying the methods of dimensionality identification. Each of $Y_t(\cdot)$ is multiplied by a weight that controls its contribution to time $t$. Hence, a model where $\delta$ functionals like model (\ref{E:bathia_model}) are aggregated can be represented as
\begin{align*}
\mathcal{Y}_t(x) = \mathcal{X}_t(x) + \mathcal{E}_t(x), \quad x\in I,
\end{align*}
where $\mathcal{Y}_t(x) \!=\! \sum_{s=t-\delta+1}^{t}\omega_{t-s}Y_s(x)$, $\mathcal{X}_t(x) \!=\! \sum_{s=t-\delta+1}^{t}\omega_{t-s}X_s(x)$ and $\mathcal{E}_t(x) \!=\! \sum_{s=t-\delta+1}^{t}\omega_{t-s}\varepsilon_s(x)$, for a positive integer $\delta$ and some coefficients $\omega_{\delta-1},\ldots,\omega_0$ so that the problem is similar to a moving average process. Denoting $\sigma^2_{\varepsilon}(x,y)=\cov(\varepsilon_t(x),\varepsilon_t(y))$, we have for $k\in\na$
\begin{align}
\cov(\mathcal{E}_t(x),\mathcal{E}_{t+k}(y)) \!=\! \left\{
\begin{array}{ccc}
0 & \text{if} & \delta-1<k,\\
\!\!\sigma^2_{\varepsilon}(x,y)\!\!\!\!\displaystyle{\sum_{s=t-\delta+1}^{t}\sum_{v=t+k-\delta+1}^{t+k}\!\!\!\!\!\!\omega_{t-s}\omega_{t+k-v}} & \text{if} & \delta - 1\geq k.\\
\end{array}
\right.
\label{E:cov_aggreg_epsilon}
\end{align}

Using the decomposition of $X_t(\cdot)$ given by (\ref{E:ProcessSpecDecomp}), we have
\begin{align}
Y_t(x) = \mu(x) + \sum_{j=1}^{d}\xi_{tj}\varphi_j(x) + \varepsilon_t(x).
\label{E:decomp_obs_func}
\end{align}
The aggregate observed function has the following decomposition:
\begin{align*}
\mathcal{Y}_t(x) &= \mu(x)\sum_{s=t-\delta+1}^{t}\omega_{t-s} + \sum_{j=1}^{d}\varphi_j(x)G_{tj}(x) + \mathcal{E}_t(x),
\end{align*}
where $G_{tj}(x) = \sum_{s=t-\delta+1}^{t}\omega_{t-s}\xi_{sj}$. It follows that
\begin{align*}
\cov&\{\mathcal{Y}_t(u),\mathcal{Y}_{t+k}(v)\} \\
&= \sum_{s=t-\delta+1}^{t}\sum_{l=t+k-\delta+1}^{t+k}\omega_{t-s}\omega_{t+k-l}M_{|l-s|}(u,v) + \cov\{\mathcal{E}_t(u),\mathcal{E}_{t+k}(v)\}.\\
\end{align*}

If $k>\delta-1$, using Equation (\ref{E:cov_aggreg_epsilon}) we have
\begin{align*}
\cov\{\mathcal{Y}_t(u),\mathcal{Y}_{t+k}(v)\} = \sum_{s=t-\delta+1}^{t}\sum_{l=t+k-\delta+1}^{t+k}\omega_{t-s}\omega_{t+k-l}M_{l-s}(u,v) = \mathcal{M}_k(u,v).
\end{align*}

Let $\lambda_1 \geq \lambda_2 \geq \ldots \geq 0$ be eigenvalues and $\varphi_1, \varphi_2,\ldots$ corresponding eigenfunctions of the operator $M_0(u,v)$. Then
\begin{align*}
\int_I M_0(u,v)\varphi_j(v)dv = \lambda_j\varphi_j(u), \quad j\geq 1.
\end{align*}
Hence, $Y_t(\cdot)$ has representation (\ref{E:decomp_obs_func}) and its serial dependence is determined by $\vxi_t = (\xi_{t1},\ldots,\xi_{td})^{\top}$, with $\e(\vxi_t)=\textbf{0}$ and $\var(\vxi_t)=\mathrm{diag}\{\lambda_1,\ldots,\lambda_d\}$. We define
\begin{align*}
\mathcal{N}_k(u,v) &= \int_I\mathcal{M}_k(u,z)\mathcal{M}_k(v,z)dz \\
&= \sum_{s}\sum_{l}\sum_{s'}\sum_{l'}\omega_{t-s}\omega_{t+k-l}\omega_{t-s'}\omega_{t+k-l'}\int_I M_{l-s}(u,z)M_{l'-s'}(v,z)dz,
\end{align*}
where $t-\delta+1\leq s,s'\leq t$ and $t+k-\delta+1\leq l,l'\leq t+k$. Here and throughout this section we shall write such summation this way, its limits being implicit.

We have that $M_k(u,v) = \sum_{i,j=1}^{d}\sigma_{ij}^{(k)}\varphi_i(u)\varphi_j(v)$, where $\Sigma_k = \e(\vxi_t\vxi^{\top}_{t+k}) = \{\sigma_{ij}^{(k)}\}$. Therefore
\begin{align*}
\mathcal{M}_k(u,v) = \sum_{i,j=1}^{d}\alpha_{ij}^{k}\varphi_i(u)\varphi_j(v),
\end{align*}
where 
\begin{align}
\alpha_{ij}^{k} = \sum_{s=t-\delta+1}^t\sum_{l=t+k-\delta+1}^{t+k}\omega_{t-s}\omega_{t+k-l}\sigma_{ij}^{(l-s)}
= \sum_{s=0}^{\delta-1}\sum_{l=0}^{\delta-1}\omega_{s}\omega_{l}\sigma_{ij}^{(l-s+k)}.
\label{E:alpha_ij_k_aggreg}
\end{align}
Since $1\leq l-s+k\leq p$, we have that $1\leq -\delta+1+k$ and $\delta-1+k\leq p$, thus $\delta \leq k \leq p-\delta+1$. Moreover,
\begin{align*}
\mathcal{N}_k(u,v) = \sum_{i,j=1}^{d}\left(\sum_{l=1}^{d}\alpha_{il}^{k}\alpha_{jl}^{k}\right)\varphi_i(u)\varphi_j(v).
\end{align*}
Then, we shall consider the operator $\mathcal{K}(u,v)=\sum_{k=\delta}^{p-\delta+1}\mathcal{N}_k(u,v)$ to estimate the process' dimension, with	 fixed integers $\delta$ and $p$, $p\geq 2\delta-1$. For the aggregate data case we consider as estimator of the covariance function
\begin{align*}
\hat{M}_k(u,v) = \frac{1}{n-\delta-p+1}\sum_{j=1}^{n-p-\delta+1}\{Y_j(u)-\bar{Y}(u)\}\{Y_{j+k}(v)-\bar{Y}(v)\},
\end{align*}
its aggregate version being given by
\begin{align*}
\hat{\mathcal{M}}_k(u,v) = \sum_{s=t-\delta+1}^{t}\sum_{l=t+k-\delta+1}^{t+k}\omega_{t-s}\omega_{t+k-l}\hat{M}_{l-s}(u,v).
\end{align*}
Therefore,
\begin{align*}
\hat{\mathcal{K}}(u,v) &= \sum_{k=\delta}^{p-\delta+1} \int_I\hat{\mathcal{M}}_k(u,z)\hat{\mathcal{M}}_k(v,z)dz \\
 &= \sum_{k=\delta}^{p-\delta+1}\sum_{s,l,s',l'}\frac{\omega_{t-s}\omega_{t+k-l}\omega_{t-s'}\omega_{t+k-l'}}{(n-p-\delta+1)^2}\sum_{i,j=1}^{n-p-\delta+1}\{Y_i(u)-\bar{Y}(u)\}\\
&\times \{Y_j(v)-\bar{Y}(v)\}\langle Y_{i+l-s}-\bar{Y},Y_{j+l'-s'}-\bar{Y}\rangle.
\end{align*}

Proposition \ref{Prop_Prop2Bathia_aggreg} shows that $\hat{\mathcal{K}}$ shares the same non-zero eigenvalues as a $(n-p-\delta+1)\times(n-p-\delta+1)$ matrix, say $\mathbf{K}^*$. Moreover, letting $\vgamma_j=(\gamma_{1j},\ldots,\gamma_{n-p-\delta+1,j})^{\top}$, $j=1,\ldots,\hat{d}$, be eigenvectors of $\mathbf{K}^*$ corresponding to the $\hat{d}$ largest eigenvalues, we have that
\begin{align*}
\sum_{i=1}^{n-p-\delta+1}\gamma_{ij}\{Y_i(\cdot) - \bar{Y}(\cdot)\}, \quad j=1,\ldots,\hat{d}
\end{align*}
are eigenfunctions of $\hat{\mathcal{K}}$. These eigenfunctions can be transformed into an orthonormal system $\hat{\psi}_1(\cdot),\ldots,\hat{\psi}_{\hat{d}}(\cdot)$ using a Gram-Schmidt algorithm.

Wavelets can be applied to aggregate data analogously to what was done in Section \ref{Sec:WaveletFuncDim}. Taking the wavelet decomposition of $Y_t(\cdot)$ on the expression for $\hat{\mathcal{K}}(\cdot,\cdot)$ we have

\begin{align*}
\hat{\mathcal{K}}(u,v) &= \sum_{k=\delta}^{p-\delta+1}\sum_{s,l,s',l'}\frac{\omega_{t-s}\omega_{t+k-l}\omega_{t-s'}\omega_{t+k-l'}}{(n-p-\delta+1)^2}\\
&\times \sum_{i,j=1}^{n-p-\delta+1} \sum_{r',r'',r'''}c_{r'}^{i}c_{r''}^{j}c_{r'''}^{i+l-s}c_{r'''}^{j+l'-s'}\phi_{r'}(u)\phi_{r''}(v), \\
\end{align*}
and considering $h_m(y) = \sum_q b_q^m \phi_q(y)$ the wavelet decomposition of an eigenfunction of $\hat{\mathcal{K}}(\cdot,\cdot)$, we have that
\begin{align*}
\int \hat{\mathcal{K}}(x,y)h_m(y)dy = \Phi(x)^{\top} \left(D\vb^m\right)\\
\end{align*}
where in this case, the element $(r',r'')$ of $D$ is given by
\begin{align*}
D_{r',r''} = \sum_{k=\delta}^{p-\delta+1}\sum_{s,l,s',l'}\!\!\frac{\omega_{t-s}\omega_{t+k-l}\omega_{t-s'}\omega_{t+k-l'}}{(n-p-\delta+1)^2}\!\sum_{i,j=1}^{n-p-\delta+1}\sum_{r'''}c_{r'}^{i}c_{r''}^{j}c_{r'''}^{i+l-s}c_{r'''}^{j+l'-s'}.
\end{align*}
Hence, we can estimate the eigenvalues of $\mathcal{K}(\cdot,\cdot)$ computing the eigenvalues of this matrix $D$, and the corresponding eigenvectors $\vb_1,\vb_2, \ldots$ contain the wavelet coefficients of the eigenfunctions of that operator.

\subsection{Density time series}

A common functional analyzed in applications is the density function of some random variable of interest.  The problem of estimating the dimension of density functions was investigated by \cite{horta2018dynamics}, which applies the method of \cite{bathia2010identifying} to financial data. The former considered as curves of interest density functions $f_t$ taking values on $L^2(I)$, $I\subset\re$. The observed densities can be taken as curves $g_t$ obtained after applying some density estimation method to the data at hand. Therefore, the assumption on the model is like Equation (\ref{E:bathia_model}), say
\begin{align*}
g_t(x) = f_t(x) + \epsilon_t(x), \quad x\in I,
\end{align*}
with $\epsilon_t$ being a noise satisfying the same assumptions made for model (\ref{E:bathia_model}), but with the additional condition that $\int \epsilon_t(x)ds = 0$, since both $f_t$ and $g_t$ must integrate one.

Wavelet based estimator for dependent time series density functions have some results established in the literature regarding its performance and consistency, like the contributions of \cite{masry1994probability, masry1997multivariate} and d Chac\'{o}n and Rodr\'{i}guez-Casal (2005). Another approach to analyze the dimension of the density functions follows from the idea of \cite{pinheiro1997estimating}, where, instead of estimating the density directly, we estimate its square root, with a wavelet estimator $\check{g}_t$ say. This change has two main advantages. First, the density can be estimated taking the square of $\check{g}_t$, which ensures that we obtain only non-negative values for the estimated density; second, letting $\check{g}_t(x) = \sum_j \check{a}_j\phi_j(x)$, $x\in I$, be the estimator's wavelet decomposition, by normalizing these coefficients such that $\sum_j \check{a}_j^2 = 1$, it follows from Parseval's identity that 
\begin{align*}
\norm{\check{g}_t}_{L^2(I)}^2 = \int_I \check{g}_t(x)^2dx = 1,
\end{align*}
which guarantees that $\check{g}_t^2$ is a \textit{bona fide} estimator of the density function and that $\check{g}_t$ belongs to $L^2(I)$. Hence, by shifting attention to $\sqrt{f_t}$ and applying the method of \cite{pinheiro1997estimating}, one can evaluate the dimension of $\sqrt{f_t}$, with the benefits of having automatically integral equal to one and non-negative estimates for the observed densities.

\section{Theoretical results}
\label{Sec:theoretical}

We prove in this section that the dimension estimators for both aggregate and non-aggregate data have the same asymptotic properties proved by \cite{bathia2010identifying} for non-aggregate data. Proposition 1 states that the eigenfunctions of $\mathcal{K}$ span the space that generates the time series functionals. Proposition 2 shows how to obtain eigenvalues of $\hat{\mathcal{K}}(\cdot,\cdot)$ as presented with the previous heuristic. The last result is a theorem showing convergence for the covariance operators and eigenvalues.

\begin{prop}
Let $\mathscr{M}$ be the space of dimension $d$ that generates the time series curves. Also, suppose $\Sigma_k = \{\sigma_{ij}^{(k)}\}$ has full rank for some $k\in\na$. Then, $\mathcal{N}_k$ and $\mathcal{K}$ (for $p\geq k$) have exactly $d$ non zero eigenvalues and $\mathscr{M}$ is spanned by the corresponding functions.
\label{Prop_Prop1Bathia_aggreg}

\begin{proof}

Denoting the adjoint operator of $\mathcal{M}_k$ by $\mathcal{M}_k^*$ \citep[Appendix A]{bathia2010identifying}, we have that $\mathcal{N}_k = \mathcal{M}_k \mathcal{M}_k^*$, since for any $f\in L^2(I)$,
\begin{align*}
(\mathcal{N}_k f)(u) &= \int_I \mathcal{N}_k(u,v)f(v)dv = (\mathcal{M}_k \mathcal{M}_k^*f)(u),
\end{align*}
where $\mathcal{M}_k^*$ is the adjoint operator of $\mathcal{M}_k$. In this case $\hat{\mathcal{K}} = \sum_{k=\delta}^{p-\delta+1}\mathcal{M}_k \mathcal{M}_k^*$. We also have $Im(\mathcal{N}_k)=Im(\mathcal{M}_k \mathcal{M}_k^*)=Im(\mathcal{M}_k)$, where $Im(\cdot)$ is the operator's image space.

We can also represent $\mathcal{M}_k$ as $\sum_{i,j=1}^{d}\alpha_{ik}^{(k)}\varphi_i\otimes\varphi_j$, then
\begin{align*}
(\mathcal{M}_k f)(u) = \sum_{i=1}^{d}\lambda_i^{(k)}\langle\varphi_i,f\rangle\rho_i^{(k)}(u),
\end{align*}
where
\begin{align*}
\rho_i^{(k)}(u) = \sum_{j=1}^{d}\frac{\alpha_{ij}^{(k)}}{\lambda_i^{(k)}}\varphi_j(u) 
\quad\text{and}\quad
\lambda_i^{(k)} = \norm{\sum_{j=1}^{d}\alpha_{ij}^{(k)}\varphi_j}.
\end{align*}
Let $\vbeta$ be an arbitrary vector in $\re^d$, $\pmb{\varphi}=(\varphi_1,\ldots,\varphi_d)^{\top}$, $\pmb{\rho}_k = (\rho_1^{(k)},\ldots,\rho_d^{(k)})^{\top}$ and $\mathbf{A}_k=\{\alpha_{ij}^{(k)}\}$, then since $\{\varphi_j,1\leq j\leq d\}$ is an orthonormal system,
\begin{align*}
\vbeta^{\top}\pmb{\rho}_k = \vbeta^{\top}\mathbf{A}_k\pmb{\varphi} = 0
\end{align*}
has a nontrivial solution iff $\vbeta^{\top}\mathbf{A}_k = \mathbf{0}^{\top}$, i.e., for all $j$
\begin{align*}
\sum_{r=1}^d\!\beta_r\alpha_{rj}^{(k)} \!=\! \sum_{r=1}^d\!\beta_r\!\!\left( \sum_{s,l}\omega_{t-s}\omega_{t+k-l}\sigma_{rj}^{(l-s)} \!\!\right)
\!\!=\! \sum_{s,l}\!\omega_{t-s}\omega_{t+k-l}\!\!\left( \sum_{r=1}^d\beta_r\sigma_{rj}^{(l-s)} \!\!\right) \!\!=\! 0,
\end{align*}
or in matrix form
\begin{align}
\begin{small}
\left[\begin{array}{c}
\omega_{\delta-1}\\
\vdots\\
\omega_0
\end{array}\right]^{\top}
\left[\begin{array}{cccc}
\sum_{r=1}^d\beta_r\sigma_{rj}^{(k)} & \sum_{r=1}^d\beta_r\sigma_{rj}^{(k+1)} & \cdots & \sum_{r=1}^d\beta_r\sigma_{rj}^{(k+\delta-1)} \\
\sum_{r=1}^d\beta_r\sigma_{rj}^{(k-1)} & \sum_{r=1}^d\beta_r\sigma_{rj}^{(k)} & \cdots & \sum_{r=1}^d\beta_r\sigma_{rj}^{(k+\delta-2)} \\
\vdots & \vdots & \ddots & \vdots \\
\sum_{r=1}^d\beta_r\sigma_{rj}^{(k-\delta+1)} & \sum_{r=1}^d\beta_r\sigma_{rj}^{(k-\delta+2)} & \cdots & \sum_{r=1}^d\beta_r\sigma_{rj}^{(k)} \\
\end{array}\right]
\left[\begin{array}{c}
\omega_{\delta-1}\\
\vdots\\
\omega_0
\end{array}\right] = 0.
\label{E:QuadFormProp1}
\end{small}
\end{align}
We know that some quadratic form satisfies $x^{\top}Ax=0$ for all $x$ iff $A$ is skew-symmetric, i.e., $A+A^{\top}$ exists and equals the zero matrix. Since the coefficients $\pmb{\omega} = (\omega_{\delta-1},\ldots,\omega_0)$ are arbitrary, this holds for all $\vomega\in\re^d$ and for all $j\in\{1,\ldots,d\}$. Therefore, the matrix in (\ref{E:QuadFormProp1}) must be skew-symmetric, which leads to
\begin{align*}
\sum_{r=1}^d\beta_r\sigma_{rj}^{(k)}=0,\quad \forall j=1,\ldots,d.
\end{align*}
Since $\Sigma_k=\{\sigma_{ij}^{(k)}\}$ is of full rank by assumption, the only solution is $\vbeta=\mathbf{0}$, which implies that $\{\rho_j^{(k)}\}$ is linearly independent. Applying the same arguments of the Lemma 2 of \cite{bathia2010identifying}, the result follows.\hfill $\square$

\end{proof}

\end{prop}

\begin{prop}
The operator $\hat{\mathcal{K}}(\cdot,\cdot)$ has the same non-zero eigenvalues of the finite matrix $\mathbf{K}^*$ of dimension $(n-p-\delta+1)\times(n-p-\delta+1)$ and whose $(m,i)$-th element is
\begin{align*}
\sum_{r=1}^{n-p-\delta+1}&\sum_{k=\delta}^{p-\delta+1}\sum_{s,l,s',l'}\frac{\omega_{t-s}\omega_{t+k-l}\omega_{t-s'}\omega_{t+k-l'}}{(n-p-\delta+1)^2}\langle Y_{m+l-s}-\bar{Y},Y_{r+l'-s'}-\bar{Y}\rangle\\
&\times\langle Y_{r}-\bar{Y},Y_{i}-\bar{Y}\rangle.
\end{align*}
\label{Prop_Prop2Bathia_aggreg}

\begin{proof}

Let $\hat{\theta}_j>0$ be an eigenvalue of $\mathbf{K}^*$ and $\vgamma_j=(\gamma_{1j},\ldots,\gamma_{n-p-\delta+1,j})^{\top}$ its corresponding eigenvector. The $m$-th element of $\mathbf{K}^*\vgamma_j=\hat{\theta}_j\vgamma_j$ is
\begin{align*}
&\frac{1}{(n-p-\delta+1)^2}\sum_{i,r=1}^{n-p-\delta+1}\sum_{k=\delta}^{p-\delta+1} \sum_{s,l,s',l'}\omega_{t-s}\omega_{t+k-l}\omega_{t-s'}\omega_{t+k-l'}\\
&~~~~~~~~~\times \langle Y_{m+l-s}-\bar{Y},Y_{r+l'-s'}-\bar{Y}\rangle\langle Y_{r}-\bar{Y},Y_{i}-\bar{Y}\rangle\gamma_{ij}=\hat{\theta}_j\gamma_{mj}
\end{align*}
Consider the function $\tilde{\psi}_j(\cdot) = \sum_{i=1}^{n-p-\delta+1}\gamma_{ij}\{Y_i(\cdot)-\bar{Y}(\cdot)\}$, then
\begin{align*}
(\hat{\mathcal{K}}\tilde{\psi}_j)(u) &= \int_I \hat{\mathcal{K}}(u,v)\tilde{\psi}_j(v)dv \\
&= \sum_{k=\delta}^{p-\delta+1}\sum_{s,l,s',l'}\frac{\omega_{t-s}\omega_{t+k-l}\omega_{t-s'}\omega_{t+k-l'}}{(n-p-\delta+1)^2}\sum_{m,r=1}^{n-p-\delta+1}\{Y_m(u)-\bar{Y}(u)\}\\
&\times \langle Y_{r}-\bar{Y},\hat{\psi}_j\rangle\langle Y_{m+l-s}-\bar{Y},Y_{r+l'-s'}-\bar{Y}\rangle \\
&= \sum_{m=1}^{n-p-\delta+1}\{Y_m(u)-\bar{Y}(u)\}\gamma_{mj}\hat{\theta}_j = \hat{\theta}_j\tilde{\psi}_j(u).
\end{align*}
Therefore, $\hat{\psi}_j$ is an eigenfunction of $\hat{\mathcal{K}}$, with corresponding eigenvalue $\hat{\theta}_j$.
\hfill$\square$
\end{proof}

\end{prop}

\begin{theorem}
\label{Theo:conv_estimators_aggreg}
Suppose the following conditions are satisfied:\\
C1. $\{Y_t(\cdot)\}$ is strictly stationary and $\psi$-mixing with the mixing coefficient defined as
\begin{align*}
\psi(l) = \mathrm{sup}_{A\in\mathcal{F}_{-\infty}^{0},B\in\mathcal{F}_{l}^{\infty},P(A)P(B)>0}|1 - P(B|A)/P(B)|,
\end{align*}
where $\mathcal{F}_i^j$ is the $\sigma$-algebra generated by $Y_i(\cdot),\ldots,Y_j(\cdot)$ for any $j\geq i$. In addition, it holds that $\sum_{l=1}^{\infty}l\psi^{1/2}(l)<\infty$.\\
C2. $\e\{\int_{I}Y_t(u)^2du\}^2<\infty$.\\
C3. $\theta_1>\cdots>\theta_d>0=\theta_{d+1}=\cdots$, i.e., all the non-zero eigenvalues of $\mathcal{K}$ are different.\\
C4. $\cov\{X_s(u),\epsilon_t(v)\}=0$ for all $s,t$ and $u,v\in I$.\\

It holds that\\

\noindent(i) $\norm{\hat{\mathcal{K}}-\mathcal{K}}_{\mathcal{S}}=O_p(n^{-1/2})$.\\
(ii) For $j=1,\ldots,d$, $|\hat{\theta}_j - \theta|=O_p(n^{-1/2})$ and
\begin{align*}
\left(\int_I\{\hat{\psi}_j(u)-\psi(u)\}^2 du\right)^{1/2} = O_p(n^{-1/2}).
\end{align*}
(iii) For $j\geq d+1$, $\hat{\theta}_j=O_p(n^{-1})$.\\
(iv) Let $\{\psi_j:j\geq d+1\}$ be a complete orthonormal basis of $\mathscr{M}^{\perp}$, and put
\begin{align*}
f_j(\cdot)=\sum_{i=d+1}^{\infty}\langle\psi_i,\hat{\psi}_j\rangle\psi_i(\cdot).
\end{align*}
Then for any $j\geq d+1$,
\begin{align*}
\left(\int_I\{\sum_{i=1}^{d}\langle\psi_i,\hat{\psi}_j\rangle\psi_i(u)\}^2du\right)^{1/2}
= \left(\int_I\left\{\hat{\psi}_j(u) - f_j(u)\right\}^2du\right)^{1/2}
= O_p(n^{-1/2}).
\end{align*}

\begin{proof}

Initially, we want to show that Theorem 1(i) of \cite{bathia2010identifying} holds also for the aggregate case. Let $\mathcal{S}$ denote the space of operators with a finite Hilbert-Schmidt norm. Since $p$ and $\delta$ are fixed and finite, we set $n\equiv n-p-\delta+1$. Let now $Z_{tk} = (Y_t-\mu)\otimes(Y_{t+k}-\mu)\in\mathcal{S}$ and consider the kernel $\rho:\mathcal{S}\times\mathcal{S}\rightarrow\mathcal{S}$ given by $\rho(A,B)=AB^*$, with $A,B\in\mathcal{S}$. We have that
\begin{align*}
\hat{M}_k\hat{M}_k^* = \frac{1}{n^2}\sum_{i=1}^{n}\sum_{j=1}^{n}\rho(Z_{ik}, Z_{ik})
= \frac{1}{n^2}\sum_{i=1}^{n}\sum_{j=1}^{n}Z_{ik}Z_{ik}^*.
\end{align*}
Therefore
\begin{align*}
\hat{\mathcal{M}}_k\hat{\mathcal{M}}_k^* &= \left(\sum_{s,l}\omega_{t-s}\omega_{t+k-l}\hat{M}_{l-s}\right)\left(\sum_{s',l'}\omega_{t-s'}\omega_{t+k-l'}\hat{M}_{l'-s'}\right)^*\\
&= \sum_{s,l}\sum_{s',l'}\omega_{t-s}\omega_{t+k-l}\omega_{t-s'}\omega_{t+k-l'}\hat{M}_{l-s}\hat{M}_{l'-s'}^*,
\end{align*}
and similarly
\begin{align*}
\mathcal{M}_k\mathcal{M}_k^* = \sum_{s,l}\sum_{s',l'}\omega_{t-s}\omega_{t+k-l}\omega_{t-s'}\omega_{t+k-l'}M_{l-s}M_{l'-s'}^*.
\end{align*}
Hence,
\begin{align*}
\norm{\hat{\mathcal{M}}_k\hat{\mathcal{M}}_k^* \!-\! \mathcal{M}_k\mathcal{M}_k^*}_{\mathcal{S}}
\!\leq\! \sum_{s,l}\sum_{s',l'}\!\omega_{t-s}\omega_{t+k-l}\omega_{t-s'}\omega_{t+k-l'}\norm{\hat{M}_{l-s}\hat{M}_{l'-s'}^* \!-\! M_{l-s}M_{l'-s'}^*}_{\mathcal{S}}.
\end{align*}
On the other hand, we note that
\begin{align*}
\hat{M}_{l-s}\hat{M}_{l'-s'}^* = \frac{1}{n^2}\sum_{i=1}^{n}\sum_{j=1}^{n}\rho(Z_{i,l-s}, Z_{i,l'-s'})
\end{align*}
is a $\mathcal{S}-$valued von Mises functional, just as $\hat{M}_k\hat{M}_k^*$, which enables us to use Lemma 3 of \cite{bathia2010identifying} to get
\begin{align*}
\e\norm{\hat{M}_{l-s}\hat{M}_{l'-s'}^* - M_{l-s}M_{l'-s'}^*}_{\mathcal{S}}^2 = O(n^{-1}).
\end{align*}
Moreover, for some distinct indexes $a$, $b$, $s$ and $l$, from Schwarz inequality we obtain
\begin{align*}
&\e\left(\norm{\hat{M}_{a}\hat{M}_{b}^* - M_{a}M_{b}^*}_{\mathcal{S}}\norm{\hat{M}_{s}\hat{M}_{l}^* - M_{s}M_{l}^*}_{\mathcal{S}}\right)\\
&~~~~\leq \left\{ \e\norm{\hat{M}_{a}\hat{M}_{b}^* - M_{a}M_{b}^*}_{\mathcal{S}}^2 \e\norm{\hat{M}_{s}\hat{M}_{l}^* - M_{s}M_{l}^*}_{\mathcal{S}}^2 \right\}^{1/2} = O(n^{-1}).
\end{align*}
Then, since the $\omega$'s, $\delta$, $p$ and $k$ are fixed, we have that
\begin{align*}
&\e\norm{\hat{\mathcal{M}}_k\hat{\mathcal{M}}_k^* - \mathcal{M}_k\mathcal{M}_k^*}_{\mathcal{S}}^2\\
&\!\leq\! \e\!\!\left(\! \sum_{s,l}\sum_{s',l'}\omega_{t-s}\omega_{t+k-l}\omega_{t-s'}\omega_{t+k-l'}\norm{\hat{M}_{l-s}\hat{M}_{l'-s'}^* - M_{l-s}M_{l'-s'}^*}_{\mathcal{S}} \!\!\right)^2
\!\!\!=\!O(n^{-1}).
\end{align*}
Thus, $\exists n_0,n_1$ such that $n\e\norm{\hat{\mathcal{M}}_k\hat{\mathcal{M}}_k^* - \mathcal{M}_k\mathcal{M}_k^*}_{\mathcal{S}}^2 \leq n_1$, $\forall n>n_0$. Then, for some $k\geq1$ and Chebyshev inequality, it follows that $\forall \epsilon>0$,
\begin{align*}
P(n^{1/2}\norm{\hat{\mathcal{M}}_k\hat{\mathcal{M}}_k^* \!-\! \mathcal{M}_k\mathcal{M}_k^*}_{\mathcal{S}} > n_1^k) \!\leq\! \frac{\e\{n\norm{\hat{\mathcal{M}}_k\hat{\mathcal{M}}_k^* \!-\! \mathcal{M}_k\mathcal{M}_k^*}_{\mathcal{S}}^2\}}{n_1^{2k}}
\!<\! \frac{1}{n_1^{2k-1}} \!<\! \epsilon,
\end{align*}
by choosing $k$ sufficiently large. This means that $\norm{\hat{\mathcal{M}}_k\hat{\mathcal{M}}_k^* - \mathcal{M}_k\mathcal{M}_k^*}_{\mathcal{S}} = O_p(n^{-1/2})$, and as consequence
\begin{align*}
\norm{\hat{\mathcal{K}} - \mathcal{K}}_{\mathcal{S}}
\leq \sum_{k=\delta}^{p-\delta+1}\norm{\hat{\mathcal{M}}_k\hat{\mathcal{M}}_k^* - \mathcal{M}_k\mathcal{M}_k^*}_{\mathcal{S}} = O_p(n^{-1/2}).
\end{align*}
Applying the same arguments of Theorem 1(ii) of \cite{bathia2010identifying}, we observe that it also holds for the aggregate case
\begin{align*}
|\hat{\theta}_j - \theta_j| = O_p(n^{-1/2})
\quad\text{and}\quad
\norm{\hat{\psi}_j - \psi_j} = O_p(n^{-1/2}), \quad j=1,\ldots,d.
\end{align*}
Additionally, we have that
\begin{align*}
\e\norm{\hat{M}_{l-s}\hat{M}_{l'-s'}^* - \hat{M}_{l-s}M_{l'-s'}^*}_{\mathcal{S}}^2 = O(n^{-2}),
\end{align*}
which gives
\begin{align*}
\e\norm{\hat{\mathcal{M}}_k\hat{\mathcal{M}}_k^* - \mathcal{M}_k\mathcal{M}_k^*}_{\mathcal{S}}^2
=O(n^{-2}).
\end{align*}
Therefore, using the same arguments of \cite{bathia2010identifying} for $\hat{\mathcal{M}}_k$ and $\mathcal{M}_k$ instead of $\hat{M}_k$ and $M_k$, we can conclude that Theorem 1(iii) and 1(iv) also holds for the aggregate case.
\hfill$\square$
\end{proof}

\end{theorem}

\section{Simulations}
\label{Sec:simulations}

In this section we present a simulation study to evaluate the performance of the method described in Section \ref{Sec:WaveletFuncDim}, to estimate the dimension of a functional via wavelets. We used the same settings employed by \cite{bathia2010identifying}, considering as true functional and noise, respectively
\begin{align*}
X_t(x) = \sum_{l=1}^d \xi_{tl}\varphi_{l}(x) \quad\text{and}\quad
\epsilon_t(x)=\sum_{i=1}^d\frac{Z_{ti}}{2^{i-1}}\zeta_i(x), \quad x\in[0,1],
\end{align*}
where for a fixed index $l$, $\{\xi_{tl},t\geq 1\}$ is an AR$(1)$ process with coefficients $(-1)^l(0.9-0.5l/d)$, the coefficients $Z_{ti}$ are independent random variables following standard normal distribution and the functionals used are
\begin{align*}
\varphi_l(x)=\sqrt{2}\cos(\pi lx) \quad\text{and}\quad \zeta_i(x)=\sqrt{2}\sin(\pi ix).
\end{align*}
The sample sizes considered are $n\in\{100,300,600\}$ and the dimensions are $d\in\{2,4,6\}$, while the maximum lag used is $p=5$. The wavelet basis is the Daubechies with four null moments. We perform a discretization of the problem, evaluating $X_t(x)$ and $\epsilon_t(x)$ on a grid of $256$ equally spaced points $x\in[0,1]$, and then, we obtain a vector of the observed functional $Y_t(x)$ evaluated at these points. For each $t=1,\ldots,n$ the decomposition for the $2^8$ points is performed using a minimum resolution level of $5$ and maximum resolution level $7$. The simulation described is based on 1000 replicates and we consider the four bootstrap procedures described in Section \ref{Sec:WaveletFuncDim}.

\begin{table}[htp]
\centering
\caption{Proportion (\%) that each value $\hat{d}$ is selected as dimension of the process when the true dimension is $d$ for each sample size $n$}
\label{T:Dim_Select}
\begin{tabular}{c|ccc|ccc|ccc}
\hline
d & \multicolumn{3}{c}{2} & \multicolumn{3}{c}{4} & \multicolumn{3}{c}{6}\\
\hline
\backslashbox{$n$}{{\small $\hat{d}$}} & 1 & 2 & 3 & 3 & 4 & 5 & 5 & 6 & 7 \\
\hline
\multicolumn{10}{c}{ordinary bootstrap}\\
\hline
100 & 46.1 & 51.8 & 2  & 41.7 & 27.9 & 1.7 & 23.6 & 8.9 & 0.4 \\
300 & 0.3 & 93.5 & 5.9 & 2.3 & 92.9 & 4.6  & 9.2 & 84.3 & 6 \\
600 & 0 & 93.3 & 6.5   & 0 & 93.2 & 6.5    & 0 & 95 & 4.9 \\
\hline
\multicolumn{10}{c}{applying thresholding before bootstrap}\\
\hline
100 & 45.4 & 51.6 & 3  & 41.6 & 28 & 1.5  & 24.3 & 11 & 0.4 \\
300 & 1.2 & 92.8 & 5.5 & 2.6 & 91.7 & 5.5 & 8 & 86.7 & 4.6 \\
600 & 0 & 94.2 & 5.8   & 0 & 94.2 & 5.4   & 0 & 95.2 & 4.7 \\
\hline
\multicolumn{10}{c}{bootstrap with the residual $\hat{\epsilon}^{thr}_t$}\\
\hline
100 & 43.1 & 53.3 & 3.6 & 42.2 & 27.7 & 1.9 & 24.9 & 8.7 & 0.1 \\
300 & 0.7 & 93.2 & 5.8  & 2.4 & 91 & 6.4    & 7.2 & 86.7 & 5.6 \\
600 & 0 & 94.6 & 5.2    & 0 & 94.3 & 5.6    & 0 & 94.9 & 5 \\
\hline
\multicolumn{10}{c}{wavestrap}\\
\hline
100 & 45.3 & 51.8 & 2.8 & 43.4 & 26.9 & 1.4 & 24.9 & 10.5 & 0.3 \\
300 & 0.9 & 93.2 & 5.6  & 2.3 & 91.4 & 6.1  & 7.5 & 88.8 & 3.4 \\
600 & 0 & 93.6 & 6.1    & 0 & 93.5 & 6.2    & 0 & 93.8 & 6.1 \\
\hline
\end{tabular}
\end{table}

\begin{figure}[htp]
\centering
\includegraphics[width=\textwidth]{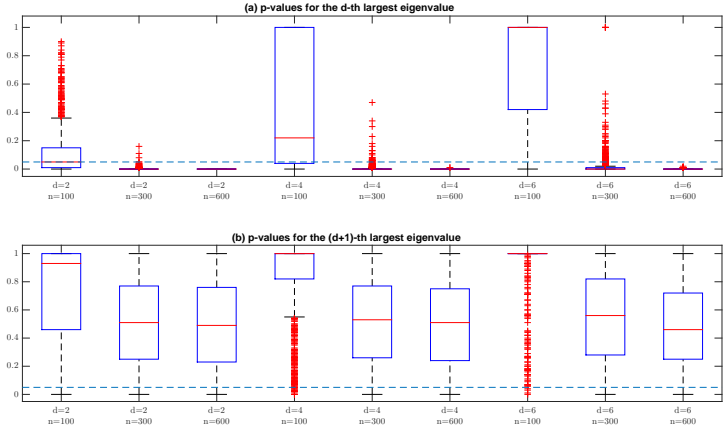}
\caption{Boxplots of the p-values of the tests for the $d$-th and $(d+1)$-th largest eigenvalues, for each sample size and true value of $d$ using the ordinary bootstrap. The segmented line represents the significance level used ($5\%$).}
\label{F:pValues_wavestrap}
\end{figure}

\begin{figure}[htp]
\centering
\includegraphics[width=\textwidth]{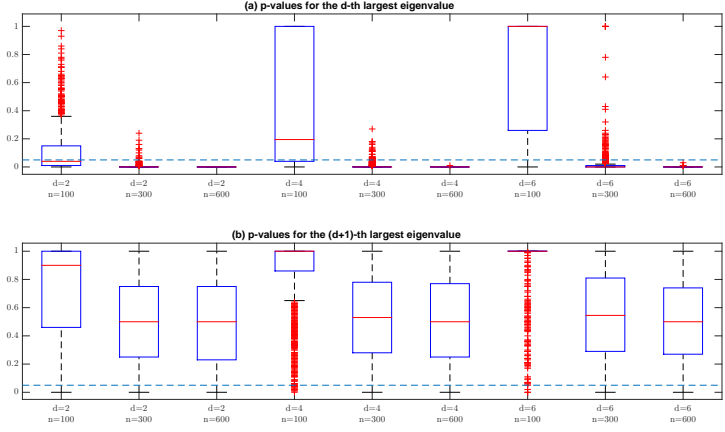}
\caption{Boxplots of the p-values of the tests for the $d$-th and $(d+1)$-th largest eigenvalues, for each sample size and true value of $d$ when the thresholding is applied before the bootstrap. The segmented line represents the significance level used ($5\%$).}
\label{F:pValues_thresh_before_wavestrap}
\end{figure}

\begin{figure}[htp]
\centering
\includegraphics[width=\textwidth]{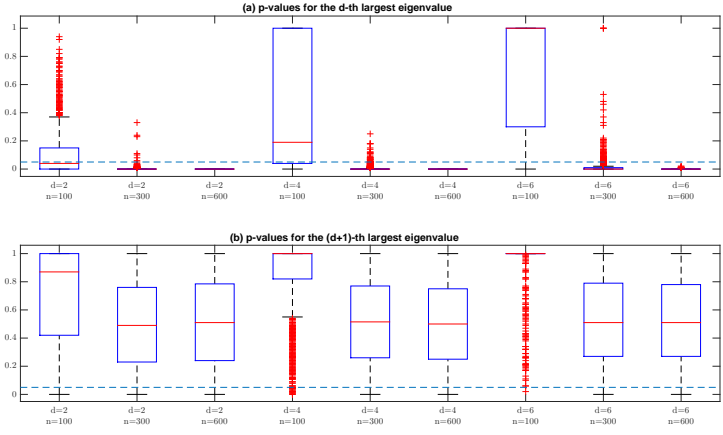}
\caption{Boxplots of the p-values of the tests for the $d$-th and $(d+1)$-th largest eigenvalues, for each sample size and true value of $d$ using the bootstrap with the residual $\hat{\epsilon}^{thr}_t$. The segmented line represents the significance level used ($5\%$).}
\label{F:pValues_residual_thresh}
\end{figure}

\begin{figure}[htp]
\centering
\includegraphics[width=\textwidth]{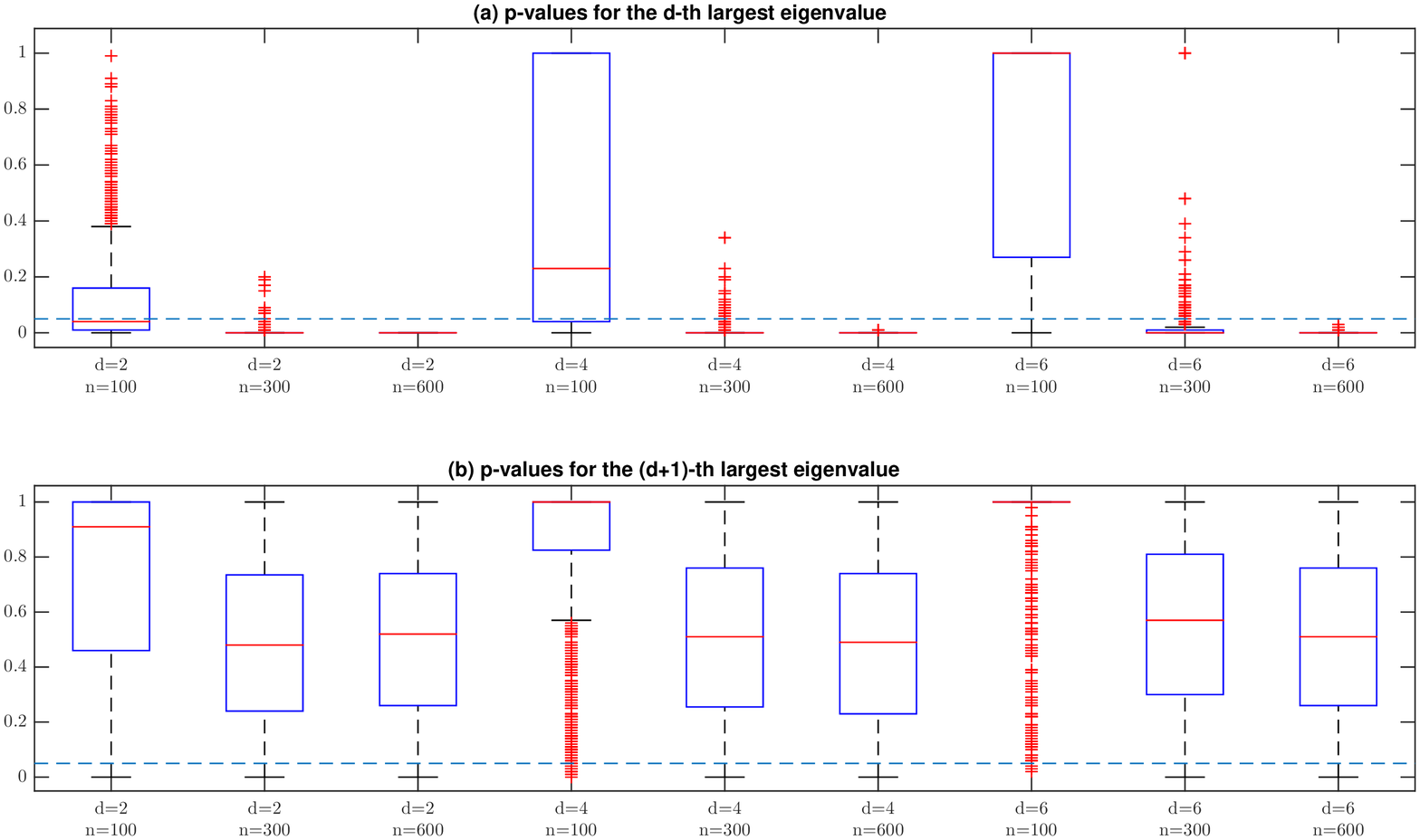}
\caption{Boxplots of the p-values of the tests for the $d$-th and $(d+1)$-th largest eigenvalues, for each sample size and true value of $d$ using the bootstrap with a wavestrapping technique. The segmented line represents the significance level used ($5\%$).}
\label{F:pValues_residual_wavestrapPSD}
\end{figure}

Table \ref{T:Dim_Select} contains the proportion each value is selected as dimension of process for different true dimensions and sample sizes. All bootstrap procedures tend to select the true value $d$ as the sample sizes increases. When the sample size is 300 or larger the four methods perform well, but have poor performance when $n=100$, especially when the true dimension has a large value. The results of the four bootstrap procedures are very close, but slight advantages can be noted for the procedures where thresholding is applied or when the residual $\hat{\epsilon}^{thr}_t$ is used. For instance, when $d=2$ these two procedures have lower rates of dimension overestimation for sample sizes 300 and 600. Figures \ref{F:pValues_wavestrap}-\ref{F:pValues_residual_wavestrapPSD} display boxplots of the bootstrap p-values for the four methods. The results in the four figures are similar, as expected while, for samples of size $n=100$, the tests select lower dimensions with considerable frequencies (not rejecting that the $d$-th largest eigenvalue is zero), but for larger sample sizes the tests present a better performance, selecting the correct dimension more often (rejecting that the $d$-th largest eigenvalue is null and not rejecting that the $(d+1)$-th largest eigenvalue is zero).

\subsection{Eigenvalues estimators for aggregate data}

In this subsection we report a simulation study performed to evaluate the eigenvalues obtained applying the method of data aggregation on the same functions used in the previous numerical evaluation. A comparison with the eigenvalues obtained without data aggregation is also presented.

The functions $\{\varphi_i,i=1,\ldots,d\}$ of the previous simulation study form an orthonormal system in $L^2([0,1])$ and the AR$(1)$ processes $\{\xi_{tl},t\geq 1\}$ have coefficients $\vartheta_l = (-1)^l(0.9-0.5l/d)$ and are independent for different $l$'s. The white noise in the AR processes are random variables with distribution $N(0,\sigma_w^2)$, where $\sigma_w^2=1.5$ was used during simulations. Therefore, the covariance function can be written as
\begin{align*}
M_k(u,v) &= \cov\{X_t(u),X_{t+k}(v)\} 
= \cov\left\{\sum_{j=1}^d\xi_{t,j}\varphi_j(u),\sum_{i=1}^d\xi_{t+k,i}\varphi_i(v)\right\} \\
&= \sum_{j=1}^d\varphi_j(u)\varphi_j(v)\sigma_{jj}^{(k)},
\end{align*}
where $\sigma_{jj}^{(k)} = \sigma_w^2\vartheta_j^k/(1-\vartheta_j^2)$, $j=1,\ldots,d$, are eigenvalues of $M_k$. Since $M_k=\sum_{i,j=1}^d\sigma_{ij}^{(k)}\varphi_i\varphi_j$ with $\Sigma_k=\{\sigma_{ij}^{(k)}\}=\mathrm{diag}\{\sigma_{11}^{(k)},\ldots,\sigma_{dd}^{(k)}\}$, for $N_k=\sum_{i,j=1}^d w_{ij}^{(k)}\varphi_i\varphi_j$ we have $W_k=\{w_{ij}^{(k)}\}=\Sigma_k\Sigma_k^{\top}=\mathrm{diag}\{(\sigma_{11}^{(k)})^2,\ldots,(\sigma_{dd}^{(k)})^2\}=\mathrm{diag}\{w_{11}^{(k)},\ldots,w_{dd}^{(k)}\}$. Thus
\begin{align*}
\int_0^1 K(u,v)\varphi_j(v)dv &= \int_0^1\left(\sum_{k=1}^p N_k(u,v)\right)\varphi_j(v)dv
= \left( \sum_{k=1}^p (\sigma_{jj}^{(k)})^2\right) \varphi_j(u),
\end{align*}
giving the eigenvalues of the function $K$ for the non-aggregate case. Applying data aggregation, the term in Equation (\ref{E:alpha_ij_k_aggreg}) is $\alpha_{ij}^{(k)}=0$ if $i\neq j$. Then $\mathcal{N}_k(u,v)=\sum_{i=1}^d (\alpha_{ii}^{(k)})^2 \varphi_i(u)\varphi_i(v)$, and
\begin{align*}
\int_0^1 \mathcal{K}(u,v)\varphi_j(v)dv 
= \left(\sum_{k=\delta}^{p-\delta+1} (\alpha_{jj}^{(k)})^2\right)\varphi_j(u),
\end{align*}
which gives the eigenvalues of $\mathcal{K}$. With these results we can compare the eigenvalues obtained with and without data aggregation with their respective true eigenvalues of the functional $X_t$.

\begin{figure}[htp]
\centering
\includegraphics[width=\textwidth]{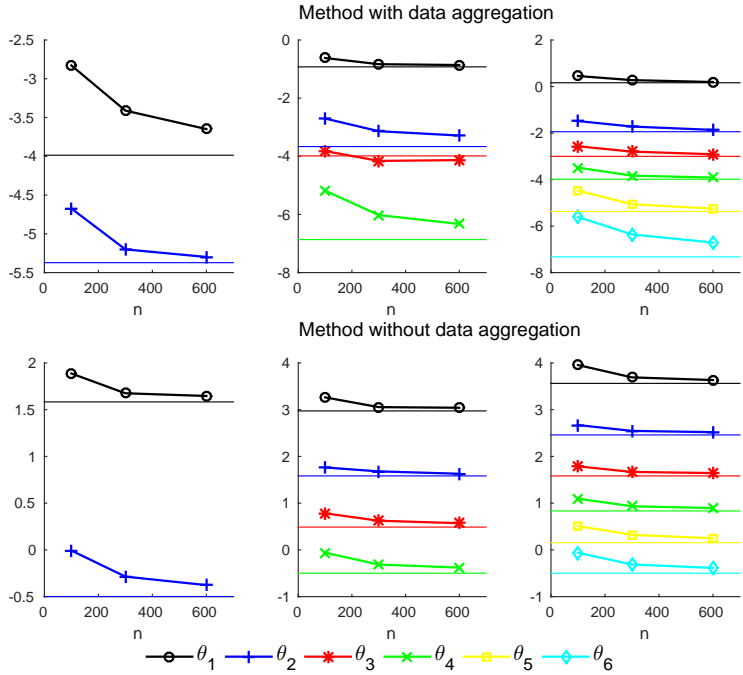}
\caption{Simulation study for aggregate and non-aggregate data. Logarithm of averages of the estimated nonzero eigenvalues (vertical axis) for different sample sizes $n$ (horizontal axis) for the methods with and without data aggregation. The number of curves in each plot corresponds to the true dimensions and horizontal solid lines correspond to the true values of the estimated eigenvalues of same color.}
\label{F:Fig_Sim_EigValAggreg}
\end{figure}

The sample sizes considered in this simulation are $n\in\{100,300,600\}$ and the dimensions are $d\in\{2,4,6\}$, while the maximum lag used is $p=5$. The numerical study is based on 1000 replicates and Table \ref{T:Sim_EigVal} presents the average of the largest eigenvalues obtained for each $n$ and $d$, as well as the true eigenvalues for each dimension. For each replicate was used $\delta=3$, with weights $\omega_2 = 0.1$, $\omega_1 = 0.3$ and $\omega_0 = 0.5$. For means of comparison, in Table \ref{T:Sim_EigVal} are also presented the analogous results applying the method of \cite{bathia2010identifying} directly, without data aggregation, which corresponds to using a single weight $\omega_0=1$. Results of Table \ref{T:Sim_EigVal} are summarized on Figure \ref{F:Fig_Sim_EigValAggreg}, which shows the estimate's averages of nonzero eigenvalues obtained for each sample size. To make it easier to discriminate between different curves, we considered the logarithm of these averages on Figure \ref{F:Fig_Sim_EigValAggreg}. Overall, we note that estimates tend to get closer to their corresponding true values as the sample size increases.

\begin{table}[htp]
\caption{Simulation study for aggregate and non-aggregate data. Average of the ten largest eigenvalues estimated for different sample sizes $n$ and true dimensions $d$ and true values of the eigenvalues for each dimension, presenting results with and without data aggregation}
\label{T:Sim_EigVal}
\begin{tabular}{c@{\hspace{\tabcolsep}}c@{\hspace{\tabcolsep}}l@{\hspace{\tabcolsep}}l@{\hspace{\tabcolsep}}l@{\hspace{\tabcolsep}}l@{\hspace{\tabcolsep}}l@{\hspace{\tabcolsep}}l@{\hspace{\tabcolsep}}l@{\hspace{\tabcolsep}}l@{\hspace{\tabcolsep}}l@{\hspace{\tabcolsep}}l}
\hline\noalign{\smallskip}
$d$ & $n$ & $\hat{\theta}_1$ & $\hat{\theta}_2$ & $\hat{\theta}_3$ & $\hat{\theta}_4$ & $\hat{\theta}_5$ & $\hat{\theta}_6$ & $\hat{\theta}_7$ & $\hat{\theta}_8$ & $\hat{\theta}_9$ & $\hat{\theta}_{10}$\\
\hline
\hline
& & \multicolumn{10}{c}{Using method with data aggregation} \\
\hline
\hline
\multirow{3}{*}{2} & 100 & 0.0591 & 0.0093 & 0.0013 & 0.0001 &  0.0000 &  0.0000 &  0.0000 &  0.0000 &  0.0000 &  0.0000\\
& 300 &  0.033 & 0.0055 & 0.0005 &  0.0000 &  0.0000 &  0.0000 &  0.0000 &  0.0000 &  0.0000 &  0.0000\\
& 600 &  0.026 &  0.005 & 0.0003 &  0.0000 &  0.0000 &  0.0000 &  0.0000 &  0.0000 &  0.0000 &  0.0000\\
& True & 0.0186 & 0.0046 & 0 & 0 & 0 & 0 & 0 & 0 & 0 & 0\\
\hline
\multirow{3}{*}{4} & 100 & 0.5383 & 0.0672 &  0.022 & 0.0056 & 0.0007 & 0.0000 & 0.0000 & 0.0000 & 0.0000 & 0.0000\\
& 300 & 0.4354 & 0.0434 & 0.0156 & 0.0024 & 0.0003 & 0.0000 & 0.0000 & 0.0000 & 0.0000 & 0.0000\\
& 600 & 0.4229 & 0.0374 & 0.0161 & 0.0018 & 0.0002 & 0.0000 & 0.0000 & 0.0000 & 0.0000 & 0.0000\\
& True & 0.3971  &  0.0256  &  0.0186 &   0.0010 & 0 & 0 & 0 & 0 & 0 & 0\\
\hline
\multirow{3}{*}{6} & 100 & 1.5702 & 0.2265 & 0.0767 & 0.0306 & 0.0114 & 0.0037 & 0.0005 & 0.0000 & 0.0000 & 0.0000\\
& 300 &  1.317 & 0.1789 & 0.0609 & 0.0217 & 0.0063 & 0.0017 & 0.0002 & 0.0000 & 0.0000 & 0.0000\\
& 600 & 1.2134 & 0.1549 & 0.0544 &   0.02 & 0.0053 & 0.0012 & 0.0002 & 0.0000 & 0.0000 & 0.0000\\
& True & 1.1759  &  0.1433  &  0.0497 &   0.0186  &  0.0046 &   0.0007 & 0 & 0 & 0 & 0\\
\hline
\hline
& & \multicolumn{10}{c}{Using method without data aggregation} \\
\hline
\hline
 \multirow{3}{*}{2} & 100 & 6.6003 & 0.9931 & 0.1515 & 0.0143 & 0.0032 & 0.0011 & 0.0003 & 0.0001 &  0.0000 &  0.0000\\
& 300 & 5.3412 & 0.7516 & 0.0553 & 0.0046 &  0.001 & 0.0004 & 0.0001 &  0.0000 &  0.0000 &  0.0000\\
& 600 & 5.1887 & 0.6861 &  0.028 & 0.0024 & 0.0006 & 0.0002 &  0.0000 &  0.0000 &  0.0000 &  0.0000\\
& True & 4.8693 & 0.6073 & 0 & 0 & 0 & 0 & 0 & 0 & 0 & 0\\
\hline
\multirow{3}{*}{4} & 100 & 26.122 & 5.8327 & 2.1884 &  0.943 & 0.2496 & 0.0034 & 0.0008 & 0.0002 & 0.0001 & 0.0000\\
& 300 & 21.212 &  5.364 & 1.8662 & 0.7305 & 0.0882 & 0.0011 & 0.0003 & 0.0001 & 0.0000 & 0.0000\\
& 600 & 20.921 & 5.0998 & 1.7707 & 0.6855 & 0.0444 & 0.0006 & 0.0001 & 0.0000 & 0.0000 & 0.0000\\
& True & 19.5567  &  4.8693  &  1.6290  &  0.6073 & 0 & 0 & 0 & 0 & 0 & 0\\
\hline
\multirow{3}{*}{6} & 100 & 52.188 & 14.454 & 5.9776 & 2.9913 & 1.6645 & 0.9401 &  0.334 &  0.001 & 0.0003 & 0.0000\\
& 300 & 40.075 & 12.757 & 5.2998 &  2.543 & 1.3797 & 0.7322 & 0.1235 & 0.0003 & 0.0001 & 0.0000\\
& 600 & 37.857 &  12.41 &  5.177 & 2.4472 & 1.2812 & 0.6805 & 0.0626 & 0.0002 & 0.0000 & 0.0000\\
& True & 35.2583 &  11.7016  &  4.8693  &  2.3012  &  1.1669  &  0.6073 & 0 & 0 & 0 & 0\\
\hline
\end{tabular}
\end{table}

\section{Application}
\label{Sec:Applic}

\begin{figure}[h]
\centering
\includegraphics[width=.9\textwidth]{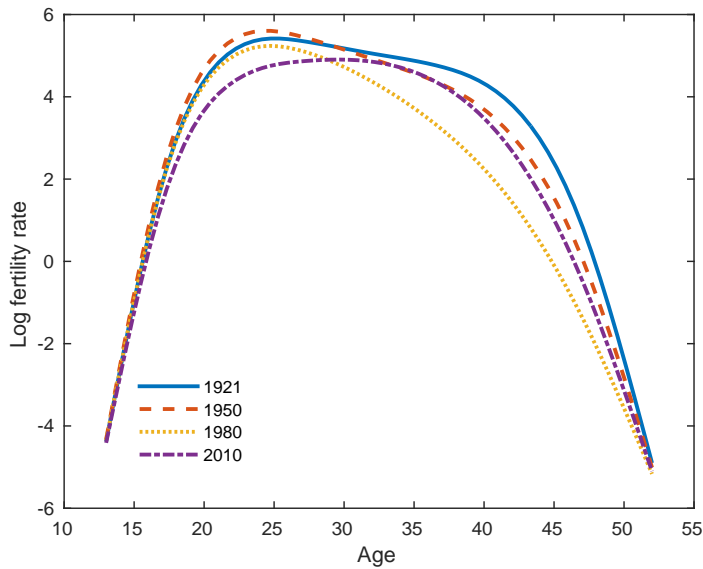}
\caption{Australian fertility rates on 1921-2010 \citep{hyndman2007robust}. The curves are logarithms of the Australian fertility rate: 1921; 1950; 1980; 2010.}
\label{F:LogRate_Age}
\end{figure}

In this section we present an application of the proposed wavelet methods in dimension estimation of functional time series. We analyze the data set of Australian fertility rates since the year 1921 until 2010. The data are in the Australian Bureau of Statistics available at \textsf{$<$http://www.abs.gov.au/AUSSTATS/} \textsf{abs@.nsf/DetailsPage/3105.0.65.0012014?OpenDocument$>$} and consist of the numbers of births per 1000 women during each year according to the age group of the mother (15-19, 20-24, 25-29, 30-34, 35-39, 40-44, 45-49). This data set was analyzed by \cite{hyndman2007robust} in the context of functional time series for the years of 1921 until 2000, where the authors considered the center of each age group as the age for which the corresponding fertility rate was observed and also assigned the value 0.005 for the ages 13 and 52 of all years. Following this procedure, we fitted a curve for the logarithm of the fertility rate of each year using smoothing splines, and considered these functionals as our observed curves. Figure \ref{F:LogRate_Age} displays the observed curves for some years. It is noteworthy from this figure that the lower log-rate for women under 25 is observed in the year 2010, which might be associated with the tendency of women in developed countries to bear less children, and later than in previous years.

To estimate the dimension of the process that generates these functionals using the wavelet based method, we evaluated the observed functionals in an equally-spaced grid of 80 points between the ages of 13 and 52. Next, we performed the procedures of Section \ref{Sec:WaveletFuncDim} using the Daubechies wavelet basis with four null moments with minimum and maximum resolution levels of 3 and 5, respectively. The value of maximum lag used was the same as in the simulation experiments, i.e., $p=5$. Table \ref{T:EigVal_FertRate} contains the five largest eigenvalues computed from the matrix $D$ obtained after the wavelet decomposition of the log-rate functionals. Table \ref{T:EigVal_FertRate} also contains eigenvalues obtained with the dimension estimation methods with and without data aggregation. For ease of comparison, all values were divided by the norm of all eigenvalues obtained with the same method. The values in Table \ref{T:EigVal_FertRate} are close for the three methods and indicate that the time series might be generated from a four- or five-dimensional process. Performing the bootstrap test for dimensionality in the wavelet based method (using the residual $\hat{\epsilon}^{thr}_t$) with 301 replications and significance level of 5\%, the result also indicates that the process has dimension 5 (we reject that $\lambda_5=0$ and fail to reject that $\lambda_6=0$).

\begin{figure}[h]
\centering
\includegraphics[width=.9\textwidth,scale=.9]{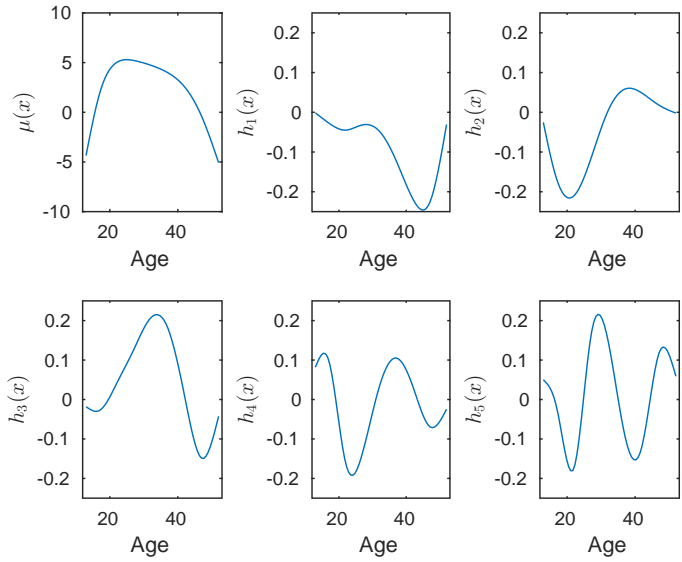}
\caption{Australian fertility rates on 1921-2010 \citep{hyndman2007robust}. Mean function and selected eigenfunctions after estimating the dimension of the process ($\hat{d} = 5$).}
\label{F:AutoFuncs}
\end{figure}

\FloatBarrier

\begin{table}[htp]
\centering
\caption{Australian fertility rates on 1921-2010 \citep{hyndman2007robust}. Five largest eigenvalues divided by the norm of all eigenvalues when applied the wavelet based method and the methods with and without data-aggregation for the fertility data}
\label{T:EigVal_FertRate}
\begin{tabular}{c|ccccc}
\hline
\multirow{2}{*}{Method} & \multicolumn{5}{c}{Eigenvalues}\\
 & $\lambda_1$ & $\lambda_2$ & $\lambda_3$ & $\lambda_4$ & $\lambda_5$ \\
\hline
Wavelet based & 99.8677 & 5.1413 & 0.0603 & 0.0021 & $<0.0001$\\
Aggregate & 99.9678 & 2.5363 & 0.0522 & 0.0003 & $<0.0001$\\
Non-aggregate & 99.9531 & 3.0614 & 0.0573 & 0.0006 & $<0.0001$\\
\hline
\end{tabular}
\end{table}

Figure \ref{F:AutoFuncs} presents the mean function and the eigenfunctions corresponding to the five largest eigenvalues. The first eigenfunction seems related to women with more than 30 years, while the second eigenfunction seems to have a relation with women below this age. The other three eigenfunctions are harder to interpret, but seem to be related to women with ages inside or outside the interval $[20,40]$, where the higher rates are usually encountered. These eigenfunctions are similar (apart from sign) to the eigenfunctions presented by \cite{hyndman2007robust} for the years of 1921 until 2000.

In their analysis, \cite{hyndman2007robust} employ a PCA with three basis function by applying procedures similar to the ones presented by Ramsay and Silverman (2005). The former authors mention that only 0.8\% of the variation is left unexplained, and from our results, this amount could still account for part (maybe not essential for their practical purposes) of the process that generates the curves. These extra information may be quite non-linear in nature.

\section{Discussion}

We study in this manuscript the problem of estimating the dimension of finite-dimensional functionals, which can be used for modeling time series of curves. This problem has been discussed by \cite{hall2006assessing} and \cite{bathia2010identifying}. The latter has used the underlying temporal stochastic structure to propose a statistical procedure which has nice asymptotic properties. We use wavelet representation in this set-up, and have attained the same asymptotic results. Moreover, besides the original bootstrap procedures, wavelets allows us to employ three additional bootstrap schemes. The wavelet method has some computational advantages as well. We also show that such method may be employed for aggregate data, and that the resulting statistical methodology has similar theoretical properties. The proposed method is illustrated in simulation studies and on a real data set.
\label{Sec:discussion}



\end{document}